# Chasing Charge Carriers: Diffusion Dynamics in Mixed-$n$ Quasi-Two-Dimensional Colloidal MAPbBr$_3$ Perovskites

Ronja Maria Piehler[a], Eugen Klein[a], Francisco M. Gómez-Campos[b], Oliver Kühn[a], Rostyslav Lesyuk[ac], Christian Klinke*[ade]

[a] *Institute of Physics, University of Rostock, Albert-Einstein-Straße 23, 18059 Rostock, Germany*

[b] *Departamento de Electrónica y Tecnología de Computadores, Facultad de Ciencias, Universidad de Granada, 18071 Granada, Spain*

[c] *Pidstryhach Institute for Applied Problems of Mechanics and Mathematics of NAS of Ukraine, Naukowa Str. 3b, 79060 Lviv, Ukraine*

[d] *Department Life, Light & Matter, University of Rostock, Albert-Einstein-Strasse 25, 18059 Rostock, Germany*

[e] *Department of Chemistry, Swansea University – Singleton Park, Swansea SA2 8PP, United Kingdom*

**Abstract**

In optoelectronic applications, metal halide perovskites (MHPs) are compelling materials because of their highly tuneable and intensely competitive optical properties. Colloidal synthesis enables the controlled formation of various morphologies of MHP nanocrystals, all with different carrier properties and, hence, different optical and carrier transport behaviours. We characterized three different methylammonium lead tribromide perovskite (MAPbBr$_3$) morphologies: nanoplatelets (NPLs), nanosheets (NSs), and nanostripes (NSTs) synthesized by hot-injection synthesis protocols with slightly different parameters. A fluorescence imaging microscope (FLIM) for time- and space-resolved measurements of the carrier migration was employed to quantify the charge carriers' migration process upon photoexcitation. The results are rationalized in the two-dimensional diffusion model framework, considering funnelling and trapping processes in mixed-$n$ colloidal MHPs. Subdiffusion mode was found to prevail in the nanocrystals, whereby the highest carrier diffusivity was found for bulk-like NSTs, followed by layered NSs and a film of NPLs. These findings provide a better understanding of optoelectronic processes in perovskites relevant to photovoltaic and light-emitting devices.

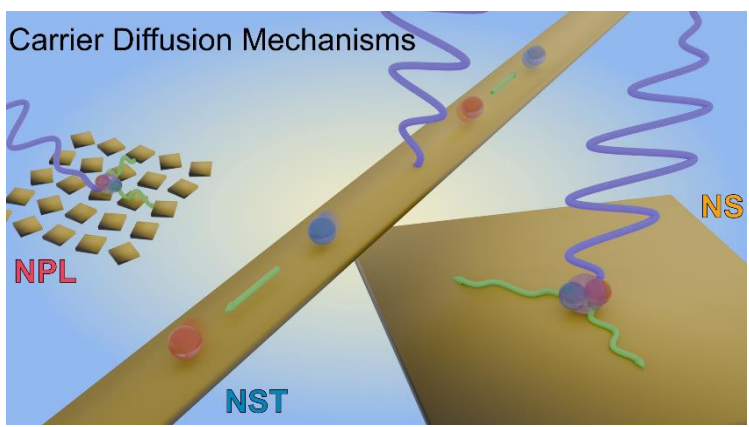

**Introduction**

Metal halide perovskites (MHP) are promising materials for the next generation of optoelectronic applications, such as light-harvesting devices, light-emitting diodes, and photodetectors.[1–4] The high potential of MHPs in optoelectronics is firmly grounded in their diverse and excellent properties, such as high defect tolerance and carrier mobility, combined with the possibility of tuneable optical properties. These properties instil confidence in the reliability and performance of these materials and pave the way for exciting advancements in optoelectronic perovskite applications.[5–8]

The general chemical formula of MHPs is $L_2[ABX_3]_{n-1}BX_4$. In the unit cell, six octahedrally arranged halogen metal atoms (e.g., bromide, iodide) represented by the X surround the divalent cation B (e.g., lead). Organic spacer cations A (e.g., methylammonium, formamidium, etc.) stabilize these metal-halide octahedra and ensure charge neutrality (Figure 1a). Surface ligands (L) isolate the MHP from the environment, building a protective shell to improve the MHP's lifetime and stability.[9,10] The ligands can also cause the MHP to have a layered structure, separating single $ABX_3$ layers from each other with thickness $n$ described by the number of $BX_3$ layers. This quasi-two-dimensionality positively affects stability and causes changes in physical properties, such as increased exciton binding energies and enhanced photoluminescence quantum yield (PLQY).[3,11] The effective band gap of the MHP shifts with increasing $n$ to lower energies.[12] Therefore, a mixed-$n$ layered MHP structure enables a funneling process, where excitons migrate to higher layer numbers with a low band gap.[12–14] The exciton binding energy increases by lowering $n$, while free charge carriers dominate bulk MHPs.[15]

Colloidal synthesis based on the hot-injection technique efficiently controls the MHP design.[16] In this report, we use colloidally prepared $(C_{12}H_{27}N)_2(MA)_{n-1}(Pb)_n(Br)_{3n+1}$ (hereinafter referred to as MAPbBr) nanoplatelets (NPLs), nanosheets (NSs), and nanostripes (NSTs) with dodecyl amine (DDA) as ligands to investigate the influence of the MHP's inner structure and shape on the diffusion behavior affected by the type of charge carriers and their recombination mechanics.

The applicability of optoelectronic perovskite devices critically depends on carrier properties, as well as their diffusion dynamics. Recent studies investigated exciton diffusion in two-dimensional MHPs, finding a subdiffusion behaviour mainly restricted by inner traps or exciton-phonon interactions.[17–19] The diffusivity was measured using a transient photoluminescence microscope.[17,18,20] It was found that dissociated free carriers cause long diffusion lengths in MHP single crystals.[21] The diffusivity of carriers within MHPs can be expressed by the diffusion coefficient $D$. We used a double-path confocal fluorescence-lifetime imaging microscope (FLIM) with a separation of the excitation and the detection path for time- and space-resolved mapping of the carrier migration, which provided input for the calculations on the diffusion coefficient $D$.

**Carrier Diffusion Measurements in Different MAPbBr Structures**

MAPbBr perovskites have long carrier diffusion lengths because of a high defect tolerance and carrier mobility.[6,7] The MHPs considered in this study are colloidally synthesized MAPbBr NPLs, NSTs, and NSs obtained in a hot-injection synthesis procedure with slightly changed parameters to obtain the different inner structures and shapes (Supplementary Information **1**).[22,23]

Figure 1b-d shows all three morphologies as transmission electron microscope (TEM) images. The NPLs have a mean lateral size of 50 nm. The edge length of the NSs varies between 2 to 5 μm. The NSTs have an average width of 35 nm with a length of up to 10 μm. The sizes along $x$- and $y$-directions in the NPLs, NSs, and NSTs are larger than the exciton Bohr radius $\simeq 3.8$ nm for bulk MAPbBr.[24] Hence, confinement effects can only arise from the inner layered structure. Compared to the NSs, the carrier diffusion in the NPLs is restricted in the $x$- and $y$-directions by the edges of the structures. The diffusion is restricted to the $x$-direction for the NSTs.

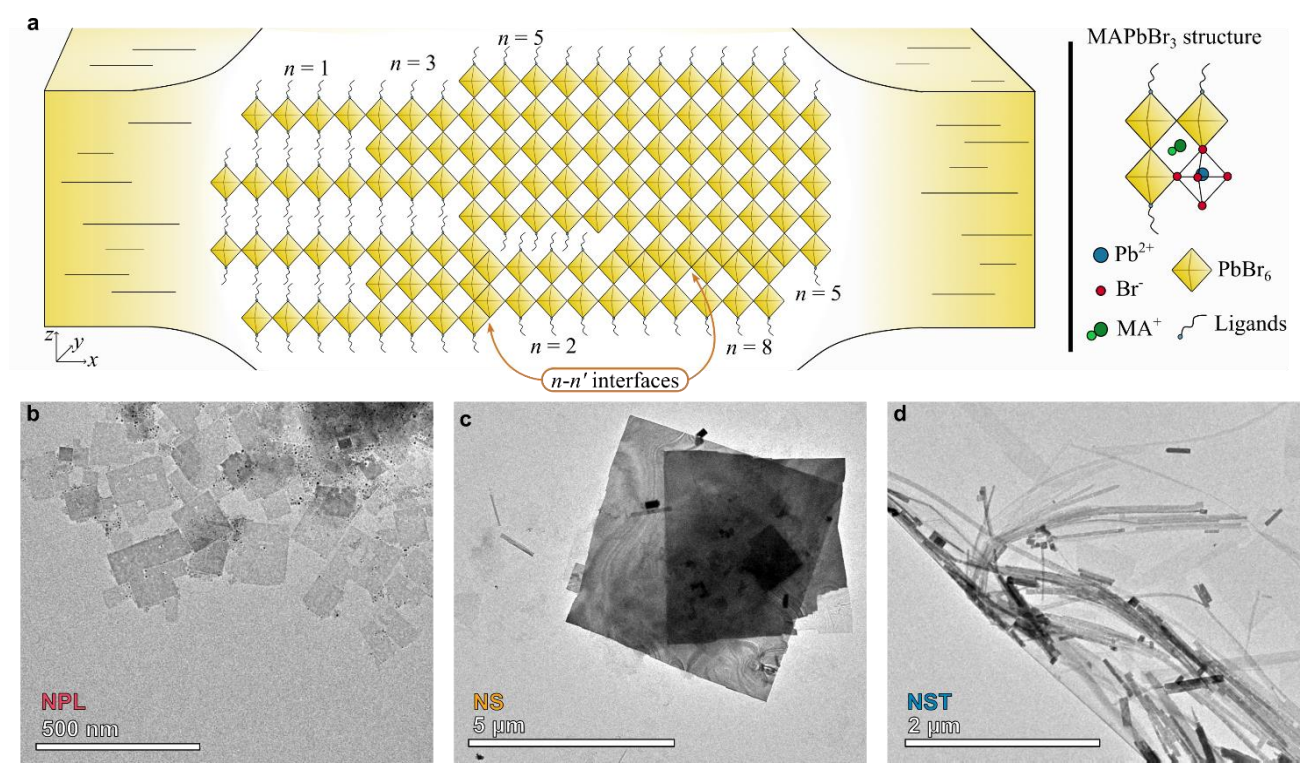

**Figure 1: Structure and Morphology – (a)** Schematic view of an internal perovskite structure containing different layer thicknesses $n = 1,2,3 ...$ and $n$-$n'$ interfaces. TEM pictures of the NPLs **(b)**, NSs **(c)**, and NSTs **(d)** synthesized in a colloidal hot-injection synthesis.

The lifetime characteristics and space- and time-resolved charge carrier diffusion are measured with a double-path confocal FLIM (Supplementary Figure 1). Figure 2a-c shows the fluorescence lifetime picture of a film of dropcasted NPLs and individual NSs and NSTs. Excitation ($\lambda = 440$ nm) and detection address the same spot. The repetition rate of the excitation pulses of 4 MHz ensures a complete decay of charge carriers. The lifetimes are visualized by the colour scale and weighted by the number of counts. The NPLs (Figure 2a) show a uniform intensity, only reduced in areas where there are no NPLs, and a lifetime of $\tau_{av} = 26$ ns at an excitation fluence of $\approx 1.06$ μJ/cm². over the whole field of view.

The intensity distribution over the single NS (Figure 2b) is not uniform, indicating differences within the structure, which will be discussed later. The lifetime is $\tau_{av} = 11$ ns at an excitation intensity of $\approx 2.12$ μJ/cm².

The NST (Figure 2c) has a uniform lifetime of $\tau_{av} = 5.83$ ns at a laser fluence of $\approx 10.61$ μJ/cm². This higher excitation fluence was necessary because of the low quantum yield from a single NST.

The corresponding diffusion maps are shown in the lower row of Figure 2 with the same **Fehler! Verweisquelle konnte nicht gefunden werden.**field of view as for the lifetime measurements. The excitation spot is fixed and in the centre of the blue spot. From this point, the movement of the charge carrier distributions can be tracked. Here, the time colour scale contains the carrier lifetime and diffusion time and is optimized to reveal a better view. The intensity weakens far from the excitation centre because long lifetime carriers diffuse further but is pronounced here for better visibility. While the diffusion of the charge carriers in the NPLs and NSs is quasi uniform in the lateral direction, appearing circular on the map, the NST reveals the bidirectional charge carrier migration manifested in elliptical intensity distribution.

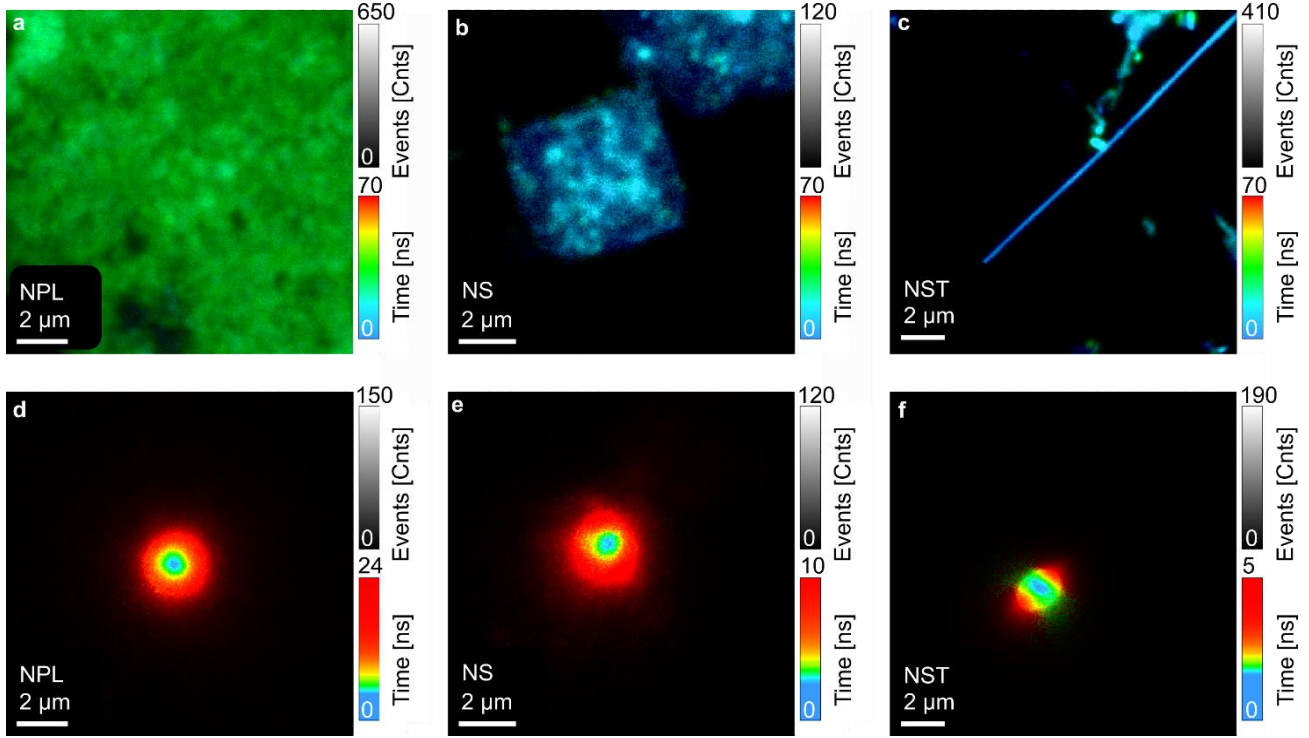

**Figure 2: FLIM lifetime and diffusion measurements** – The upper row shows spatially resolved images of the carrier lifetime weight by intensity. The dropcasted NPL film **(a)** shows an isotropic intensity distribution and the same lifetime in the whole field of view. For the NS **(b)**, bright spots, meaning higher intensity, can be seen with a consistent lifetime. The lifetime is the same over the whole NST **(c)** as the intensity. **(d-f)** The fixed excitation (blue) and separated detection path enable spatial and time-resolved measurements of the charge carrier diffusion. The timescale indicates the carrier diffusion time to its recombination spot. A nearly isotropic diffusion can be seen for the NPL film **(d)** and the NS **(e)**. The NST shows the diffusion along its shape **(f)**.

FLIM measurements enable dividing pictures into consecutive time steps. Because of a low intensity, single time frames are binned to images of 2 ns. A Gaussian distribution fits the forward diffusion broadening.[17,18,25] Isotropic diffusion is assumed for the NS and NPL, meaning equal broadening in the $x$ and $y$-direction. Therefore, the variance $\sigma_x^2(t) = \sigma_y^2(t)$ represents the broadening over time $t$. The NST also show a bidirectional diffusion but is strongly limited in one direction by the edge of the NST (Supplementary equations (1), (2)).

Figure 3 shows the fitted distribution of radiatively recombined carriers for NPLs, NS, and NST after different times $t$ (raw data can be found in Supplementary Information **4**). The broadening of the Gaussians is clearly visible. The intensity decreases over time due to the migration concentric from the excitation spot and recombination of the charge carriers. For further clarity, the insets show the current broadening and diffusion distance determined by the difference between subsequent slices (Supplementary Information **5**). An intensity < 0 (blue) represents recombined carriers, whereas leaving carriers by further diffusion appear in the regions where the intensity is > 0 (red). Time lap videos visualize this diffusion and can be found in the Supplementary Information.

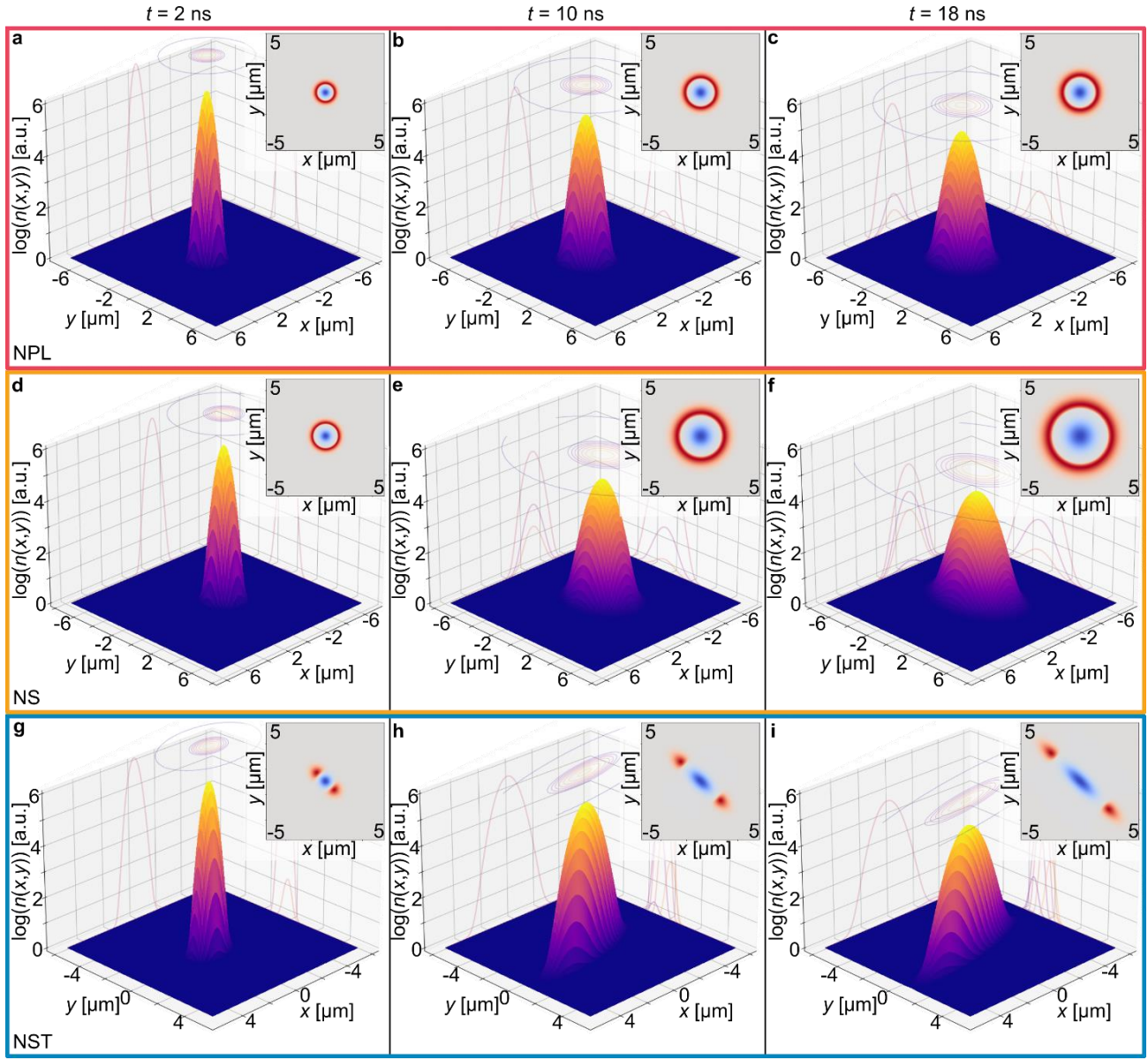

**Figure 3: Carrier diffusion** – The measured distribution of charge carriers within a film of NPLs **(a-c)**, an NS **(d-f)**, and an NST **(g-i)** are fitted with a Gaussian and broaden in $x$ and $y$-direction from $t = 2$ ns to $t = 18$ ns. The logarithm of the intensity of the distribution is represented for visibility. The according time lap videos visualize the broadening over time and can be found in the Supplementary Information. The raw data can be found in the Supplementary Information **4**. The insets show two subtracted consecutive time steps. The intensity is normalized to the range of $-1$ (blue) to $1$ (red) and represents the change of leaving ($< 0$) and arriving ($> 0$) carriers. In the area in between, recombining and diffusing carriers are in balance (0, grey). The outer grey area is also 0 since there are no charge carriers.

In Figure 3, differences in diffusion become apparent when comparing NPLs, NSs, and NSTs. Accordingly, NSTs show the longest diffusion length, followed by NSs and NPLs. In this study, the charge carrier dynamics in MHPs will be analyzed quantitatively using the general diffusion equation.[17] With this, the measured intensity of the two-dimensional exciton density distribution $n(x,y,t)$ for monomolecular and bimolecular recombination can be described.

$$\dot{n}(x,y,t) = G(x,y,t) + D\nabla^2 n(x,y,t) - k_1 n(x,y,t) - k_2 n(x,y,t)^2 \qquad (1)$$

with $G(x,y,t)$ as the particle (charge carrier or excitons) generation term. For $t = 0$, the initial laser excitation is a Gaussian with a width of 245 nm according to the resolution limit and the given geometries by the setup. For $t > 0$ $G(x,y,t) = 0$ applies. The monomolecular or bimolecular recombination rates are $k_1$, representing recombination due to excitons or trapped charge carriers, and $k_2$, recombination of free carriers, respectively. The diffusion is determined by the diffusion coefficient $D$.

To apply this equation to the measured NPLs, NS, and NSTs carrier diffusion and to determine the differences in the diffusion behaviour, further knowledge about the MHPs is needed. Therefore, additional measurements were done to disclose the inner structure, type of carriers, and recombination mechanisms.

**Characterization of the MAPbBr Structures**

X-ray diffraction (XRD) was performed to characterize the inner structure. Figure 4a-c shows strong reflections from the MAPbBr single-crystal (100) and (200) directions for all three investigated MHP morphologies.[26] Some peaks are not visible due to the texture effect because of the planar orientation of the nanocrystals on the substrate. In the NPLs and NSs, additional peaks below (100) and (200) are visible and assigned to the mixed-$n$ layered structure of the MHPs with layers of $n = 2,3, ... \infty$, where $n = \infty$ describes a bulk structure.[22,26] The contribution from higher $n$ layers is more pronounced in the NSs. In the NPLs, artifacts from the layered structure are also visible. The fact that these peaks are broader than those of the NS is due to the fact that the NPLs are thin and hence not all layered but single layers meaning low $n$.[27] The NSTs are composed entirely of bulk crystal structures.

The type of carriers was examined by steady-state ensemble absorption and photoluminescence (PL) measurements in solution. The absorption and the PL spectra are normalized to their maximum and shown in Figures 4d and e, respectively. The central emission peak is at $\lambda_{em} = 515$ nm for all three structures. Therefore, this peak must be assigned to a bulk structure with high $n$ since it is the lowest energy gap for the carriers to recombine, and the NSTs are bulk only. These results suggest carrier funnelling from regions with low $n$ (higher energy states) to regions with higher $n$ (lower energy states) in the layered NSs and NPLs.[12,22] Since the DDA ligands used in the synthesis are short (12 carbon chain length compared to 14 (tetradecylamine) or 16 (hexadecylamine)), they support the funnelling process.[22] The NSs also show smaller shoulders at shorter wavelengths. Here, the radiative recombination of the charge carriers happens before relaxation to the lowest energy, which is a sign of separated layer domains with low $n$. These shoulders are not dominant in the NPL and NST emissions, indicating the possibility of complete relaxation to the bulk and recombination at low energy.

As mentioned in the lifetime picture, the intensity distribution over the single NS (Figure 2b) is not uniform. The number and combination of mixed layers between the distinguishable bright and dark regions can differ.

Since the excitation wavelength is at $\lambda = 440$ nm, electron-hole pairs can be generated only in layers where the optical gap is below $\approx 2.81$ eV, which could lead to a disadvantage in excitation for regions with low $n$ which have a larger bandgap (2.91 eV for $n = 3$ for NSs) as discussed below and therefore showing low PL intensity there. The lattice structure of MAPbBr is considered as mechanically soft leading to strong electron-phonon interactions which are even more pronounced in layered perovskites.[28–30] This energy transfer to phonons can explain the reduced PL intensity at spots where $n$ is low.

In the absorption spectrum (Figure 4d), strongly pronounced exciton peaks above the bulk band gap energy $E_g$ are visible for the NPLs and NSs. Since every layer $n$ has its own band gap energy and continuum states these absorption peaks support the layered structure from the XRD measurements. Since there is more than one peak in the absorbance spectra, different MHPs can have different thicknesses or a mixed-$n$ layered structure. The positions of the absorption peaks fit with previous studies and can be assigned to layers of $n = 2$ to $n = 8$, while higher layers are supposed to be bulk structures (Supplementary Information **6**).[14,27,31] Since those peaks are not pronounced in the absorption spectrum of the NSTs, they are again confirmed to have predominantly a bulk-like structure.

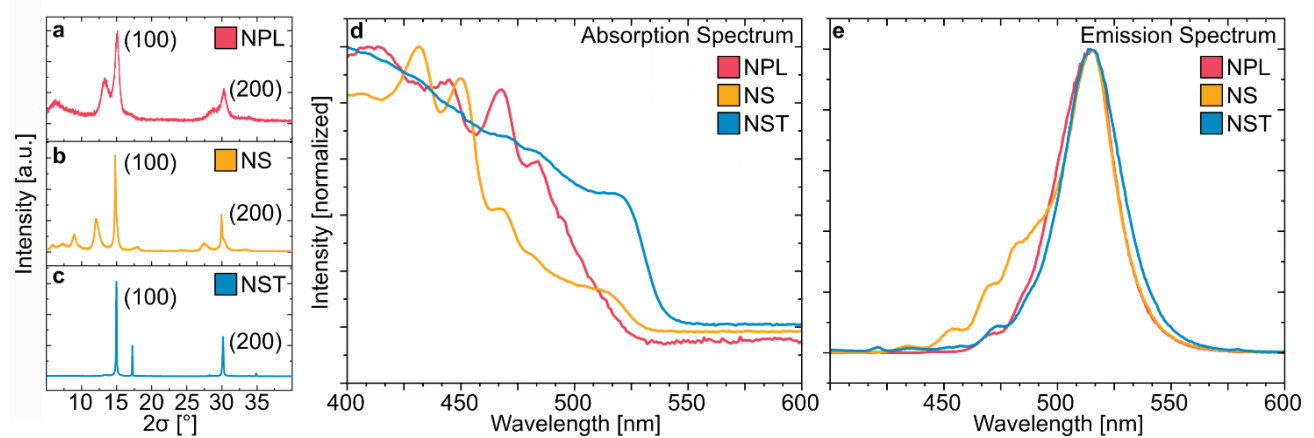

**Figure 4: Different MHP structures – (a-c)** XRD patterns show peaks for the (100) and (200) directions of the MAPbBr crystal structure for all three samples measured in solution. NPLs and NSs have additional equidistant peaks, which can be assigned to different layers $n$. Steady-state ensemble absorption **(d)** measurements show exciton peaks for lower wavelengths for the NPLs and NSs, indicating a mixed-$n$ layered structure, while the NSTs only have bulk contributions. **(e)** The central PL peak at 515 nm is the same for all samples. Small shoulders for the NSs show emission in low $n$ layers, while the emission in NPLs and NSTs is dominated by bulk.

A multiple Elliott model fitting helps to extract the exciton binding energies $E_b$ and band gap energies $E_g$ of the different layer components and is explained in the Supplementary Information **7**.[32–34]

The NSs exhibit four exciton peaks for energies above the bulk band gap. However, the maximum exciton binding energy for the NSs is $E_b = 98.5$ meV for the highest measured exciton peak with a band gap energy of $E_g = 2.97$ eV.[14,31] Because of this high $E_b$, the dominant carriers in the NSs are assumed to be excitons. The maximum binding energy for the NPLs is $E_b = 156.4$ meV for the peak with a band gap of $E_g = 2.81$ eV [14,31] Since there is practically only bulk for the NSTs, only the contribution at the lowest energy band gap was fitted, leading to $E_g = 2.38$ eV and $E_b = 19.9$ meV, which is still below the room temperature $E = k_b T \approx 25$ meV. Hence, excitons will dissociate to free charge carriers and are expected to be the dominant species in the NSTs for all excitation energies at room temperature, in contrast to NPLs and NSs where excitons are formed in the dominate exciton peaks of low $n$.[14,31]

Measuring the fluorescence lifetime with a FLIM enables a more detailed examination of the charge carrier dynamics and recombination mechanisms of individual MHP structures. The post synthesis solution containing the MHPs was diluted and drop-casted onto glass slides to identify individual structures or, in the case of the NPLs, to have a film. The results of three representative individual structures for each the NSs and NSTs, or in the case of the NPLs three different positions on the film are averaged for a more general statement. Figure 5a-c shows the lifetime measured for different laser fluence $P_{\text{laser}}$ for NPLs, NSs, and NSTs. The lifetime curves are fitted with a triexponential tail fit (Supplementary Information **8**). All nanostructures have a similar average lifetime $\tau_{av}$. It decreases with increasing laser fluence (Figure 5a) due to the higher number of excitons or charge carriers and associated faster recombination.[16]

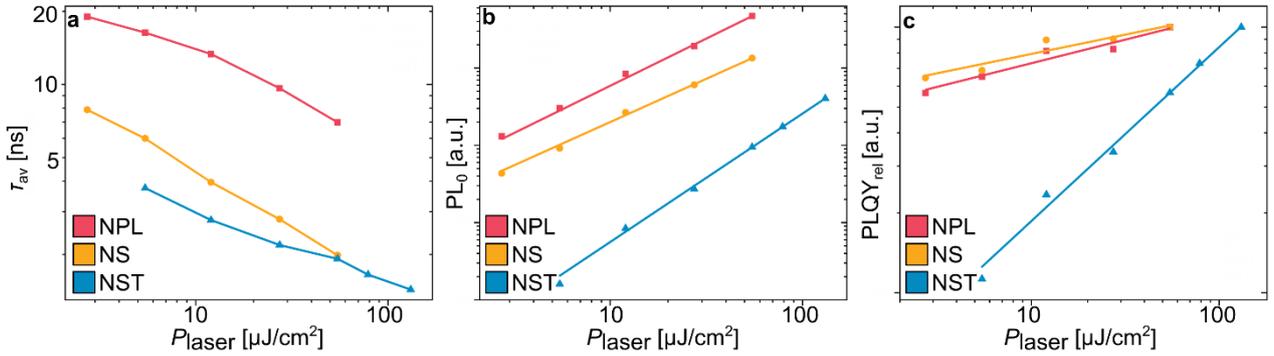

**Figure 5: Lifetime measurements** – **(a)** The average lifetime $\tau_{av}$ is plotted over the excitation power $P_{\text{laser}}$. **(b)** The intensity of the decay curves $PL_0$ at $t = 0$ is fitted with a power law to analyze the present recombination mechanisms and, hence, the charge carriers. Monomolecular recombination dominates for NPLs and NS, while the NSTs are closer to bimolecular recombination; these assumptions are supported by the rel. PLQY in **(c)**. The measurements were done for 3 samples for each morphology.

The intensity $PL_0$ at the maximum of the decay curves is plotted over the excitation fluence (Figure 5b). Fitting with the power law $PL_0 = a \cdot P_{\text{laser}}^b$ and analyzing the slope $b$ allows to analyze the present recombination mechanisms. The NPLs and NSs have a slope of $b = 1.22 \pm 0.05$ and $b = 1.120 \pm 0.025$, respectively. A slope of $b \approx 1$ indicates a monomolecular recombination process as found for excitons or trapped charge carriers. Funnelling in the mixed-n layered MHPs leads to the dissociation of the excitons since the exciton binding energy decreases with increasing layer number $n$ and is faster or comparable to trapping in defects.[14,35,36] Free charge carriers would show a bimolecular recombination behaviour, where $b \approx 2$. Hence, the funnel paths occur as quasi-traps at the $n$-$n'$-interfaces. This explains the slope as being slightly higher than one. The NSTs show $b = 1.662 \pm 0.013$ and are closer to a bimolecular recombination process indicating free charge carriers.[37,38] The slope $b < 2$ can be considered as a fingerprint of defect-trapped charge carriers recombining in a monomolecular mode.

The division of $PL_0$ by the laser fluence can be determined as the relative $PLQY_{\text{rel}}$, which also characterizes the present charge carrier dynamics (Figure 5c). Again, the power law can be used for fitting $PLQY_{\text{rel}} = A \cdot P_{\text{laser}}^c$ and leads to a slope of $c = 0.178 \pm 0.026$ and $c = 0.145 \pm 0.028$ for the NPLs and NSs, respectively which can be well explained within the monomolecular regime.[39] In the NSTs with three times larger $c = 0.655 \pm 0.023$ bimolecular recombination dominates.[39]

**Carrier Diffusion Dynamics in Various MAPbBr Morphologies**

Knowledge about the inner structure, type of carriers, and recombination mechanisms in the considered NPLs, NS, and NSTs allows to differentiate the diffusion behaviour (Figure 1), which can now be characterized. For this, we numerically solved equation (1) in one dimension with the finite difference method and used a cross-section of the data as a compare function in a Python minimization method (Supplementary Information **10**). Additionally, a two-dimensional fitting for the NPLs and NSs was done to verify the numerically fitting (Supplementary Information **11**).

From the fitting we obtained the diffusion constant $D$ and the two recombination rates $k_1$ and $k_2$. The monomolecular recombination $k_1$ was in the same order of magnitude for all morphologies, while the bimolecular recombination rate $k_2$ was higher for the NSTs ($10^{-4}$) than for the NPLs and NSs ($10^{-6}$). Therefore, the monomolecular recombination mechanism predominates in the NPLs and NSs, as discussed above in the lifetime considerations, and confirms the quasi-trapping character of the funnelling. In contrast, in the NSTs where the $k_2$ is larger, bimolecular and monomolecular decay mechanisms happen side by side, which also goes hand in hand with the previous observations.

Figure 6 shows the fitting results for the diffusion constant $D$ at different laser intensities. The diffusion constant $D$ is highest for NSTs, followed by NSs and NPLs. Although the NPLs and NSs show similar structure and carrier dynamic characteristics, the diffusion constant of the latter is slightly higher. The reduced $D$ of the NPLs could be due to hopping between particles in the film hampering diffusion.[27]

Even though the NSTs cannot be assigned as a pure one-dimensional sample and carriers can diffuse in both lateral directions $x$ and $y$, the diffusion is strongly limited in one direction. For this purpose, the diffusion constant was halved to be comparable with the definitive two-dimensional diffusions of NPLs and NSs. (Supplementary Information **10**). However, the NSTs show the highest diffusion constant $D$. The free charge carriers are mainly responsible for a higher diffusion constant than NPLs and NSs. Another reason for the difference between the bulk and layered structures is the funneling paths in the latter. As discussed, the funnels could act as quasi-traps at the $n$-$n'$-interfaces, hindering carrier diffusion. The free carriers in the NSTs are almost only restricted by inner and surface defects, which are also present in the NPLs and NSs.

As was shown in previous studies, $D$ increases with increasing laser fluence for all measured MHP structures.[41] Increasing the generation rate relative to the recombination rates, hence, higher charge carrier concentration can explain this improved diffusion. The enhanced number of charge carriers can cause the traps to fill[16,42], facilitating diffusion.[17,18,41] The trapping in the NPLs and NS can be seen in the subdiffusion behavior (Supplementary Information **9**).

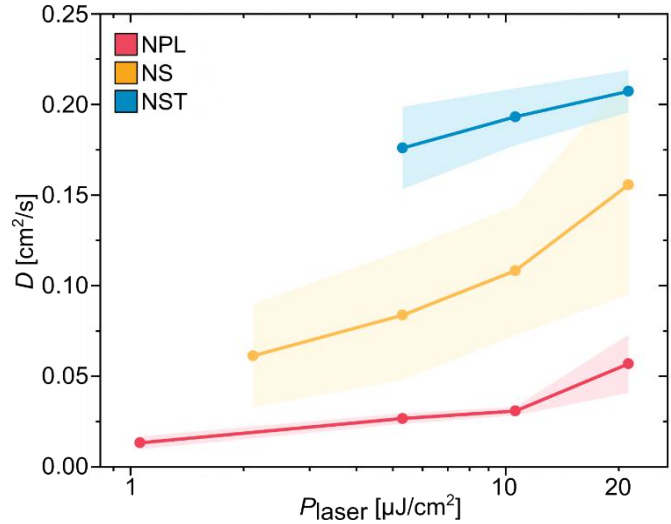

**Figure 6: Resulting diffusion behaviors of NPLs, NSs, and NSTs –** The diffusion constant $D$ increases with increasing laser fluence for all NPLs (red), NSs (yellow), and NSTs (blue). The NSTs show the highest diffusivity, followed by the NSs and NPLs. For comparison with the NPLs and NSs the one-dimensional kind diffusion of the NSTs is halved (Supplementary Information **10**). The error bars marked by the filled area show the deviation of the single measured structures from the mean value of $D$ for the three measured samples for each morphology.

**Conclusion**

In summary, we demonstrated and characterized the spatial and time-resolved charge carrier diffusion in MAPbBr NPLs, NSs, and NSTs using a double-path FLIM setup. The layered structure was first analyzed with XRD, and steady-state UV-Vis and PL spectroscopy of the NPLs and NSs and led to high exciton binding energies (≈100 meV) supporting the formation of excitons. Lifetime measurements and applying the multiple Elliott fitting of the absorbance spectra revealed excitons as the dominant species in the mixed-$n$ layered MHPs. Free charge carriers govern the charge carrier dynamics of the bulk structure of the NSTs. These differences in monomolecular recombination for the NPLs and NSs, and bimolecular recombination for the NSTs are also confirmed by the finite difference minimization fitting of the analyzed diffusion maps. The diffusion in the NSs and NPLs is reduced because of the quasi-trap character of the funnels. Moreover, in the latter, the hopping of the excitons from one NPL to the next reduces the diffusivity. The NSTs show the highest diffusivity because the free charge carriers are almost only restricted by defects and are not subject to such hindrances as interfaces between different $n$-domains and funnels in contrast to NSs and NPLs. The funnels can play a very positive role in outcompeting the trapping process in MHPs. However, they might also present a natural problem for diffusivity in layered MHPs, suggesting the necessity of compromise for the design of efficient optoelectronic devices. Overall, the colloidal MHPs with different morphologies provide a reasonable basis for observing charge carrier diffusion phenomena by the double-path confocal FLIM technique. This provides a vital understanding of charge carrier migration to improve MHP optoelectronic applications further.

**Experimental Section**

Transmission Electron Microscope (TEM) – The samples were prepared by drop-casting the diluted suspension on a TEM copper grid coated with a carbon film and measured on a *Talos-L120C* and *EM-912 Omega* at 120 kV and 100 kV.

Fluorescence Lifetime Imaging Microscope (FLIM) – The confocal microscope platform *Micro Time 200* (PicoQuant) was used. The fibre-coupled picosecond laser has an excitation wavelength of $\lambda = 440$ nm. A flippable mirror allows an easy switch between the two operation modes – fluorescence lifetime and diffusion measurements. For the diffusion measurements, the sample is excited always in the centre of the field of view. The detection path is guided pixel-wise by a galvo-scanner over the field of view with a minimal possible step size of about 10 nm. With the decoupling of the excitation and detection path, a time-correlated single photon counting (TCSPC) method measured with a single photon detector (PMA Hybrid) allows the tracking of the carriers over space and time (Supplementary Figure 1).

X-ray diffraction (XRD) – XRD patterns were measured with a *Panalytical Aeris System* with a Bragg-Brentano geometry and a copper anode with an X-ray wavelength of 0.154 nm from the Cu-kα1 line.

Spectroscopy – The absorption spectra were obtained with a *Lambda 1050+ spectrophotometer* (Perkin Elmer) equipped with an integration sphere. The PL spectra were measured with a *Spectrofluorometer FS5* (Edinburgh Instruments).

Statistical Analysis – The Gaussian functions of the carrier diffusion were fitted in *Python* with the optimization function "scipy.optimize". The spectra intensities for absorption and emission are normalized by division by the maximum in *Origin*. PL Lifetime analysis were fitted in *Origin*. For fitting the diffusion equation a finite difference method was used and fitted in *Python* with "scipy.minimize". (Supplementary Information)

**Acknowledgements**

Deutsche Forschungsgemeinschaft (DFG, German Research Foundation) is acknowledged for funding of SFB 1477 "Light-Matter Interactions at Interfaces", project number 441234705, W03 and W05. We also acknowledge the European Regional Development Fund of the European Union for funding the PL spectrometer (GHS-20-0035/P000376218) and X-ray diffractometer (GHS-20-0036/P000379642) and the Deutsche Forschungsgemeinschaft (DFG) for funding an electron microscope Jeol NeoARM TEM (INST 264/161-1 FUGG) and an electron microscope ThermoFisher Talos L120C (449942675, INST 264/188-1 FUGG).